\documentclass[12pt, a4wide]{article}
\usepackage{amssymb, amsmath}
\begin{document}

\title{Macroscopic Unobservability of Spinorial Sign Changes}
\author{Richard D.~Gill\thanks{Mathematical Institute, 
University of Leiden; gill@math.leidenuniv.nl; +31-6-224\,27\,127}}

\date{March 22, 2015}        

\maketitle

\begin{abstract}
\noindent In Section 5 of Christian (2014) an experiment is described which is purported to have the capacity for exhibiting quantum correlations in a completely classical environment. Unfortunately the experiment has an interesting self-destructive property: it is certain \emph{not} to deliver the required result. Unfortunately, this makes it pretty certain that no experimenter will ever bother to perform it. 
\end{abstract}

\noindent In Section 5 of Christian [1], an experiment is described which is purported to have the capacity for exhibiting quantum correlations in a completely classical setting. Unfortunately the experiment has an interesting self-destructive property: it is certain \emph{not} to deliver the required result, hence it is pretty certain that no experimenter will ever bother to perform it.

The experiment involves colourful exploding balls which separate into pairs of spinning hemispheres. Later in the section these are replaced by pairs of squashy balls which are initially squeezed together. With the help of a battery of video cameras (later:  three or more successive laser screens) and state of the art image processing software, $n$ angula momenta $\mathbf s ^ k$, $k = 1, \dots, n$, get stored in a data-base. Within the $k$th pair, one ``particle'' has spin angular momentum $+\mathbf s ^ k$, the other $-\mathbf s ^ k$. In fact (and as Christian remarks) we only need the direction of each real, three-dimensional vector $\mathbf s ^ k$; its length is irrelevant in the ensuing calculations.

Next, two reference directions $\mathbf a$ and $\mathbf b$ are chosen. Christian asks for them to be chosen randomly from the set of already existing observed directions; however, his aim is to experimentally determine a correlation function, $\mathcal E(\mathbf a, \mathbf b)$ in which $\mathbf a$ and $\mathbf b$ both vary throughout the unit sphere $S^2$. Presumably $n$ is large and the observed directions are spread throughout $S^2$ so the remarkable restriction that we only measure correlations for pairs of directions which have actually been observed is not much of a restriction.

The formula which he states that the experimenter has to use (i.e., omitting a hypothetical limit for $n\to\infty$) is 
$$
\mathcal E(\mathbf a, \mathbf b) ~=~ \frac 1 n \sum_{k=1}^n \text{sign}(+\mathbf s ^ k\cdot \mathbf a) 
\, \text{sign}(-\mathbf s ^ k\cdot \mathbf b).
$$
Now pick any two pairs of directions $\mathbf a^1$, $\mathbf a^2$ and $\mathbf b^1$, $\mathbf b^2$. 
Define $x^{ki} = \text{sign}(+\mathbf s ^ k\cdot \mathbf a^i)$ and $y^{kj} = \text{sign}(-\mathbf s ^ k\cdot \mathbf b^j)$, and define $\mathcal E^{ij} = 
\mathcal E(\mathbf a^i, \mathbf b^j) = n^{-1}  \sum_{k=1}^n x^{ki}y^{kj}$. It follows that 
$$\mathcal E^{11} - \mathcal E^{12} - \mathcal E^{22} - \mathcal E^{21} =  \frac1n  \sum_{k=1}^n \bigl(x^{k1}(y^{k1}-y^{k2}) - x^{k2}(y^{k2} + y^{k1})\bigr).$$ 
Since for each $k$ the four numbers $x^{k1}$, $x^{k2}$, $y^{k1}$, $y^{k2}$ are all equal to $\pm 1$, one of the two expressions $y^{k1}-y^{k2}$ and  $y^{k1}+y^{k2}$ equals $0$ and the other equals $\pm 2$. It follows that each of the $n$ terms $x^{k1}(y^{k1}-y^{k2}) - x^{k2}(y^{k1} + y^{k2})$ equals $\pm 2$ and their average lies between $-2$ and $+2$. It is therefore impossible, as is well known (CHSH inequality) for $\mathcal E(\mathbf a, \mathbf b)$ to be close to the famous singlet correlation function $-\cos(\mathbf a\cdot \mathbf b)$.

Aside from this extraordinary mistake, Section 5 of the paper reproduces the sign error present in the preprint Christian [2] and both earlier and later papers (and a book) by the same author, and exposed by Gill [3], as well as by numerous other authors. The advantage of paper [2] for those who want to study Christian's works is its brevity: it is just one page; and hence its refutation can also be rather brief. The material in [2] is incorporated without change in Christian's later book. Florin Moldoveanu (personal communication) has kindly pointed out for me where the sign error is hiding in the present attempt. Specifically, in [1], equation (110) is incorrect because it is adding two kinds of $L_{\rho}(\lambda)$ belonging to different algebras defined by equation (94) for two values of $\lambda$. Treating $L_{\rho}(+1)$ and $L_{\rho}(-1)$ on equal footing implies from equation (94) that $L_{\rho}(+1) = L_{\rho}(-1) = 0$ as one can see by subtracting equation (94) for $\lambda = 1$ from itself when $\lambda = -1$: the standardized variables $A(\mathbf a, \lambda)$ and $B(\mathbf b, \lambda)$ vanish.

Christian argues in Section 5 of [1] (as in his other papers) that correlations should be computed by taking account of bivectorial standard errors, seemingly contradicting his own instructions to the experimenters. Redefining correlation in such a complex way allows Christian plenty of space for hiding a sign error; its location has shifted about over the many papers he has written but the bump under the carpet does not go away so easily. Once the sign error is corrected, the bivector correlation is no longer a scalar but contains a nonzero bivector which has no possible interpretation, exposing in another way the folly of his bivectorial generalisation of standard probability theory. .

Bell-type experiments by design generate discrete outcomes, forcing the correlations to be computed in the standard way. Theory has to predict these observed correlations. As is well known, Bell's inequality can be rewritten as an inequality between probabilities of different combinations of discrete outcomes. The experimenter merely counts, and compares observed relative frequencies to predicted probabilities, see Weatherall [4].

\section*{References}
\raggedright
\frenchspacing
\noindent 
[1] ~ J.~Christian (2014). Macroscopic Observability of Spinorial Sign Changes under $2\pi$ Rotations. \textit{Int.\ J.\ Theor.\ Phys.} \\ DOI 10.1007/s10773-014-2412-2
\medskip

\noindent 
[2] ~  J.~Christian (2011). Disproof of Bell's Theorem.  \texttt{arXiv:1103.1879}

\medskip
\noindent 
[3] ~ R.D.~Gill, (2012). Simple refutation of Joy Christian's simple refutation of Bell's simple theorem. \texttt{arXiv:1203.1504}

\medskip
\noindent 
[4] ~ J.O. Weatherall (2013). The Scope and Generality of Bell's Theorem. \textit{Found. Phys.} \textbf{43}, 1153--1169.

\end{document}